# Toward a Science of Autonomy for Physical Systems: Transportation

Daniel Lee
ddlee@seas.upenn.edu
University of Pennsylvania

Sebastian Pokutta
sebastian.pokutta@isye.gatech.edu
Georgia Institute of Technology

Transportation systems are currently being transformed by advances in information and communication technologies. The development of autonomous transportation holds the promise of providing revolutionary improvements in speed, efficiency, safety and reliability along with concomitant benefits for society and economy. It is anticipated these changes will soon affect household activity patterns, public safety, supply chains and logistics, manufacturing, and quality of life in general.

## Impact on Society and Economy

Development of autonomous transportation systems is proceeding with breathtaking speed, and these systems will continue to progress in maturity, robustness, trustworthiness, and usability. While future projections vary drastically, most automotive companies expect vehicles with combined function automation to be a reality by 2020 and IEEE projects that by 2040 about 75% of all vehicles will be autonomous[2], with the potential of $1 trillion in annual savings and up to 1 gigaton of reduced carbon emissions due to shared, electrified, and autonomous vehicles[3]. Regardless of the precise numbers, there will undoubtedly be benefits and advantages in safety, convenience, energy, sustainability, supply chains, land use and public transportation as detailed below[4].

**Improved safety:**
Modern transportation systems engineering has successfully reduced much of the risk from technical failures. At the same time we have witnessed in recent years a shift towards the human element as a potential source of system failure. Recent high-profile examples demonstrate that these incidents could have been prevented with various levels of autonomy in the transportation systems:

1. **Germanwings Flight 9525 crash in March 2015[5]:** A pilot locked the cockpit and flew himself and 149 others into a mountain to commit suicide. Lufthansa officials (Germanwings is a subsidiary of Lufthansa) had earlier pronounced the pilot "100% fit-to-fly" after he had passed all medical evaluations, showing why it is so hard to manage and quantify risk arising from a human operator. Minutes before the crash, telemetric data from the plane clearly indicated that the aircraft was in distress, descending at a rate of 18 m/s. At this time an autonomous system could have been activated by the air traffic controller to take emergency control of the plane .

2. **Amtrak train 188 derailment in May 2015[6]:** The Amtrak train was traveling at a speed of 102 mph in a 50 mph zone, which led to derailment of the train and left 8 people dead and 200 injured. As pointed out by various officials the derailment likely would have been prevented with an autonomous speed management and train control system.

3. **Drunk driving accidents and general traffic safety:** While traffic fatalities from drunk driving have been declining, they still remain one of the prime contributors to traffic fatalities and accidents and the estimated total cost to the United States from drunk driving amounts to $199bn/year[7]. Further, currently about 2.5 million people are injured and 41,000 people are killed annually in highway accidents in the US[8], most of which are caused due to human error. Autonomous vehicles and automated driving functions could drastically mitigate the number of accidents caused by human error.

4. **Evacuation scenarios:** In the case of an evacuation of a metropolitan area, central planning and coordination is key to a safe, fast, and reliable execution. However, panic in the face of adversarial events gives rise to chaos and uncoordinated actions. Autonomous transport can play a crucial role in evacuation procedures, as they can be coordinated and stabilize evacuation flow and speed. Emergency evacuation scenarios with autonomous transportation can be efficiently managed, stabilizing overall evacuation flow and speed.

---

[1] Contact: Ann Drobnis, Director, Computing Community Consortium (202-266-2936, adrobnis@cra.org).
For the most recent version of this essay, as well as related essays, please visit: cra.org/ccc/resources/ccc-led-white-papers
[2] http://www.ieee.org/about/news/2012/5september_2_2012.html
[3] Connected and Autonomous Vehicles, 2014 Vision, PennDOT Report, Hendrickson, et. al.
[4] http://cleantechnica.com/2015/03/14/us-transportation-system-could-save-1-trillion-annually-reduce-carbon-emissions-by-1-gigaton/

---

[5] http://www.nytimes.com/2015/04/19/world/europe/germanwings-plane-crash-andreas-lubitz-lufthansa-pilot-suicide.html?_r=0
[6] http://en.wikipedia.org/wiki/2015_Philadelphia_train_derailment
[7] http://www-nrd.nhtsa.dot.gov/Pubs/812013.pdf
[8] http://www-nrd.nhtsa.dot.gov/Pubs/811016.PDF and http://www-nrd.nhtsa.dot.gov/Pubs/811017.PDF

**Convenience:**

Another obvious area of key impact is convenience and lifestyle. Many households in the United States exhibit a mobility pattern bound by the shared usage of a small number of owned vehicles. Examples include transporting children to school and other activities. Autonomous transport could drastically improve mobility for children and free time on parents' schedules. Another scenario applies to transportation services for elderly or handicapped persons who currently depend upon human help. In the above cases, general household activities would be enhanced and positively affected by autonomous transport.

Autonomous transportation also holds significant promises for commuting. Today, commuting is associated with a significant loss of time and productivity; on average Americans spend between 25-30 minutes commuting each direction. While the actual commuting time may remain constant (Marchetti's constant)[9] autonomous transportation could provide a mobile living/working space where the commuter can turn travel time into productive time. Leveraging autonomy in the context of personal transportation will lead to a significant increase in quality of life, lower costs, and regained productivity.

**Energy and Sustainability:**

In 2010 transportation accounted for about 70% of all petroleum consumption and about 27% of overall energy consumption in the US[10]. Autonomous transportation could significantly impact this consumption via much more efficient *hypermiling* as compared to human driving as well as significantly improved efficiency by higher vehicle utilization in shared systems[11]. These benefits are significantly compounded once vehicles-to-vehicle and vehicle-to-infrastructure communications complement automation. Vehicles no longer have to anticipate and predict various traffic patterns that might require aggressive actions, instead traffic intersection infrastructure could inform the approaching vehicles of optimal passage times and synchronize acceleration and deceleration. Vehicle-to-vehicle communication will also allow vehicles to platoon in a synchronized fashion, drastically reducing traffic congestion and travel times[12].

**Highly Efficient Supply Chains, Manufacturing, and Logistics:**

Manufacturing contributes about $2 trillion to the US economy, accounting for about 12% of the GDP, and supporting about 18 million jobs in the US[13]. Many manufacturing operations are built around the concepts of lean and just-in-time[14] supply chain and logistics operations in order to keep them economically feasible. This requires materials and goods to arrive at the respective facility often only

---

[9] http://en.wikipedia.org/wiki/Marchetti%27s_constant
[10] http://en.wikipedia.org/wiki/Energy_in_the_United_States
[11] http://blog.rmi.org/blog_2014_09_09_energy_implications_of_autonomous_vehicles
[12] http://machineslikeus.com/news/vehicle-communication-prevent-traffic-congestion
[13] http://www.nam.org/Newsroom/Facts-About-Manufacturing/
[14] http://en.wikipedia.org/wiki/Just-in-Time_Manufacturing

hours ahead of their actual use and disruptions and variations in the arrival of material and goods can be disastrous to the operation. Autonomous transport can mitigate many of these effects by ensuring more stable traffic patterns as well as eliminate rest times required for drivers to further improve efficiency. Currently about 69% of all goods shipped in the US are moved via long-distance trucking[15] and about 20-40% of the overall shipping cost arises from fuel. Autonomous systems can significantly reduce fuel costs by platooning, and improve overall travel time and reliability while significantly reducing overall costs.

**Land use:**
Another area that will be positively affected by autonomous transportation is land use and urban design and sprawl. One prime example in this category is parking. Autonomous vehicles do not need to be parked close to the passengers location, e.g. home or workplace, but rather could return to more remote depots until requested, or they could serve another passenger within a car-sharing setup. Moreover, fully autonomous transportation systems will require less dedicated roads and lanes freeing up high value land in urban areas[16]. If car sharing concepts using autonomous vehicles become more prevalent, individual car ownership might drastically decrease to a point that impacts private land use, e.g., land occupied by driveways and garages could be repurposed. Autonomous vehicles might also counteract recent trends of urban overcrowding by re-enabling urban sprawling[17].

**Public Transportation:**
Autonomous transportation will also impact and transform public transportation systems. More traditional modes of public transportation (buses, trains, metro, etc.) will not only be augmented with autonomous technology but will also be supplemented by customized public transportation via self-driving cars for shorter trips. Various states in the US (California, Nevada, Michigan and Florida) have already passed legislation to allow self-driving cars on the streets and many other states are debating similar bills and following suit. Traffic information and demand data can be further used to obtain optimized dynamic routes and schedules, which in turn will drastically improve efficiency of the public transportation system. The Singapore Land and Transportation Authority predicts that shared autonomous transportation will potentially reduce the total number of cars on the road to approximately 20% of today's number, leading to significant reductions in pollution, energy consumption, congestion, and travel time[18].

**Broader Societal and Economic Impact:**
While autonomous transportation may replace several jobs directly associated with transportation, it can also spur the creation of new higher value jobs. These could be related to maintaining, enabling, and operating fleets of autonomous vehicles, as well as unforeseen new industries enabled by autonomous transportation systems. Moreover autonomous transport will likely impact some fundamental assumptions of society in terms of vehicle ownership. Current utilization of vehicles is highly inefficient, with most vehicles idle approximately 90% of the time.  Autonomous transportation will enable much more efficient shared usage, and this will reduce congestion and improve sustainability and service levels.

---

[15] http://www.supplychain247.com/article/why_trucks_will_drive_themselves_before_cars_do
[16] http://www.arupconnect.com/2014/10/08/road-diets-and-car-clouds-shaping-the-driverless-city/
[17] http://www.slate.com/blogs/moneybox/2014/10/15/self_driving_tesla_car_might_encourage_urban_sprawl.html
[18] http://www.lta.gov.sg/content/ltaweb/en/publications-and-research/reports/annual-reports.html

## Challenges and Enablers

The vision outlined above will not be immediate but rather requires significant investment and research in various key areas to resolve crucial challenges on the way to realizing the benefits of autonomous transportation.  The challenges include technological problems that relate directly to autonomous transportation and transportation infrastructure, reliability and trustworthiness challenges pertaining to operating these vehicles, and regulatory challenges. We detail some of these issues below.

**Technology:**
There are a number of technological challenges that need to be solved before autonomous transportation systems can become ubiquitous.  Some examples of existing technological issues include reliable sensing, navigation, and networking.

**Sensing:** Current autonomous vehicles rely heavily on precise sensing and location information.  For example, Global Positioning Systems (GPS) are needed to accurately track the vehicle pose. However, GPS may be unavailable or inaccurate due to storms, trees or building cover and multi-path effects.  Robust vehicular tracking systems need to be developed that can handle these situations. Another key system is LIDAR[19] which uses reflected laser light to create an accurate 3D-representation of the surrounding environment. Current LIDAR systems are rather expensive, costing more than the price of the vehicle. Moreover, LIDAR sensors have limitations in snow and rain[20][21], making it difficult to construct accurate geometrical maps. To overcome these challenges, better sensors and processing algorithms need to be developed to provide more reliable and accurate estimates under adverse conditions.

**Navigation:** Human operators are very good at adjusting to unexpected changes and navigating in uncertain environments. In contrast, current autonomous systems require a variety of prior information in the form of preprocessed maps and environmental data[22].  In unexpected situations such as emergency road closures or construction, autonomous systems may not be able to successfully navigate in these conditions. The presence of other actors, such as erratic pedestrians or malicious drivers, can also make it difficult for autonomous systems to reliably determine the appropriate actions to take. Additional research is needed in these situations for autonomous driving systems to devise robust representations of the external environment and other agents and to make optimal decisions. Extensive tests will also be required to ensure the systems perform the correct and appropriate interactions.

---

[19] http://en.wikipedia.org/wiki/Lidar
[20] http://www.technologyreview.com/news/530276/hidden-obstacles-for-googles-self-driving-cars/
[21] http://www.quora.com/Can-self-driving-cars-deal-with-inclement-weather
[22] http://www.technologyreview.com/news/530276/hidden-obstacles-for-googles-self-driving-cars/

**Connected vehicles and networking:** An important component to a successful autonomous transportation system is communications, including vehicle-to-vehicle (V2V), vehicle-to-infrastructure (V2I), and vehicle-to-auxiliary (V2X) aspects. So far many efforts have been focused on the individual vehicle and these communication systems will play an important role in enabling autonomous transportation systems across larger scales by augmenting the local sensors onboard the individual vehicles. An example is when maps and environmental information from a number of vehicles are fused and shared, enabling each vehicle to access a more accurate representation of the surrounding world. However, a significant amount of research is still needed in order to identify the best ways to share and distribute this information in real-time.

**Reliability and Trustworthiness:**

**Security:** There is concern that increased autonomy in transportation systems will be more vulnerable to hacking due to their increased reliance on computer systems and electronic control units[23]. It is imperative that improved security and encryption methods be developed and deployed to minimize the risk of malicious users causing serious harm and damage to autonomous transportation systems. In addition to potential physical misuse, there is also the risk of private information from networked vehicles such as personal driving history and patterns being released. New measures and standards will be needed to protect drivers and the general public from the misuse of this technology.

**Trust:** There is also the question of how to build trust between human users and autonomous systems. Is it possible to validate and guarantee certain levels of performance in these systems? Methods need to be established to test, validate and certify the correct performance of new autonomous transportation systems, similar to product safety certification by OSHA but with rigorous standards applied to autonomous systems and behaviors.

**Regulation and Oversight:**
Finally it will be imperative to have a solid regulatory framework in place to guide liability and usage questions as well as encourage the adoption of autonomous transportation systems in a manner that benefits society. This framework has to be developed in close dialog with experts in autonomous transportation systems to appropriately reflect technological and engineering-related aspects.

---

[23] Tracking and Hacking: Security and Privacy Gaps Put American Drives at Risk, Markey Report, Feb. 2015.

## Recommendations and Conclusions

The coming implementation of autonomy and related technologies will have a major impact in the future of transportation systems. It is critical that we better understand how advances in sensors, mapping, navigation, data analytics, security and other technologies will influence safety and efficiency of transportation. A renewed commitment to studying and preparing for these upcoming changes is needed, encompassing the federal level to corporations as well as the general public.

*For citation use*: Lee D. & Pokutta S. (2015). *Toward a Science of Autonomy for Physical Systems: Transportation*: A white paper prepared for the Computing Community Consortium committee of the Computing Research Association. http://cra.org/ccc/resources/ccc-led-whitepapers/

*This material is based upon work supported by the National Science Foundation under Grant No. (1136993). Any opinions, findings, and conclusions or recommendations expressed in this material are those of the author(s) and do not necessarily reflect the views of the National Science Foundation.*